\pdfoutput=1
\documentclass[sigconf, table, rgb, natbib=false, screen, 9pt, nonacm]{acmart}

\copyrightyear{2023}
\acmYear{2023}
\setcopyright{acmlicensed}\acmConference[ISPD '23]{Proceedings of the 2023 International Symposium on Physical Design}{March 26--29, 2023}{Virtual Event, USA}
\acmBooktitle{Proceedings of the 2023 International Symposium on Physical Design (ISPD '23), March 26--29, 2023, Virtual Event, USA}
\acmPrice{15.00}
\acmDOI{10.1145/3569052.3578928}
\acmISBN{978-1-4503-9978-4/23/03}

\usepackage[utf8]{inputenc}
\usepackage[T1]{fontenc}
\usepackage[english]{babel}
\usepackage{csquotes}
\usepackage{graphicx}
\usepackage{mathtools,nicefrac,nicematrix}

\usepackage[
	backend=biber,
	style=ieee,
	sortlocale=en_US,
	sortcites=true,
	url=true,
	eprint=true,
	giveninits=false,
	dashed=false,
	minnames=1,
	maxnames=5
	]{biblatex}
\AtEveryBibitem{
  \clearname{editor}%
  \clearfield{series}%
  \clearfield{isbn}%
  \clearfield{issn}%
  \clearfield{day}%
  \clearfield{month}%
  \clearfield{place}%
  \clearlist{location}%
  \ifentrytype{software}{%
  }{%
  \clearfield{url}%
  \clearfield{urlyear}%
  }%
}

\usepackage{caption}
\usepackage{subcaption}
\usepackage{hyperref}

\DeclarePairedDelimiter\mket{\lvert}{\rangle}

\newcommand*{\ket}[1]{\ensuremath{\mket{\mkern1mu#1}}}

\usepackage{listings}
\definecolor{codegreen}{rgb}{0,0.6,0}
\definecolor{codegray}{rgb}{0.5,0.5,0.5}
\definecolor{codepurple}{rgb}{0.58,0,0.82}
\definecolor{backcolour}{rgb}{0.95,0.95,0.92}
\definecolor{tumblue}{RGB}{0,101,189}
\lstdefinestyle{mystyle}{
    backgroundcolor=\color{backcolour},
    commentstyle=\color{codegreen},
    keywordstyle=\color{tumblue},
    numberstyle=\tiny\color{codegray},
    stringstyle=\color{codepurple},
    basicstyle=\ttfamily\footnotesize,
    breakatwhitespace=false,
    breaklines=true,
    captionpos=b,
    keepspaces=true,
    numbers=left,
    numbersep=5pt,
    showspaces=false,
    showstringspaces=false,
    showtabs=false,
    tabsize=2,
    frame=lines
}

\newtheorem{example}{Example}

\emergencystretch=\maxdimen
\begin{document}
\title{MQT QMAP: Efficient Quantum Circuit Mapping}
\subtitle{(Invited Summary Paper)\vspace*{3mm}}

\author{Robert Wille}
\affiliation{%
        \institution{Chair for Design Automation\\Technical University of Munich}
		\city{Munich}
        \country{Germany}
}
\affiliation{%
        \institution{Software Competence Center Hagenberg GmbH}
		\city{Hagenberg}
        \country{Austria}
}
\email{robert.wille@tum.de}
\orcid{0000-0002-4993-7860}

\author{Lukas Burgholzer}
\affiliation{%
        \institution{Institute for Integrated Circuits\\Johannes Kepler University Linz}
		\city{Linz}
        \country{Austria}
}
\email{lukas.burgholzer@jku.at}
\orcid{0000-0003-4699-1316}
\renewcommand{\shortauthors}{Wille and Burgholzer}

\ccsdesc[500]{Hardware~Quantum computation}
\ccsdesc[500]{Hardware~Emerging languages and compilers}
\ccsdesc[300]{Hardware~Physical design (EDA)}
\ccsdesc[300]{Hardware~Hardware description languages and compilation}
\ccsdesc[100]{Software and its engineering~Compilers}

\keywords{quantum computing, quantum circuit mapping, compilation}

\begin{abstract}
Quantum computing is an emerging technology that has the potential to revolutionize fields such as cryptography, machine learning, optimization, and quantum simulation. However, a major challenge in the realization of quantum algorithms on actual machines is ensuring that the gates in a quantum circuit (i.e., corresponding operations) match the topology of a targeted architecture so that the circuit can be executed while, at the same time, the resulting costs (e.g., in terms of the number of additionally introduced gates, fidelity, etc.) are kept low. This is known as the quantum circuit mapping problem. This summary paper provides an overview of QMAP---an open-source tool that is part of the \emph{Munich Quantum Toolkit}~(MQT) and offers efficient, automated, and accessible methods for tackling this problem. To this end, the paper first briefly reviews the problem. Afterwards, it shows how QMAP can be used to efficiently map quantum circuits to quantum computing architectures from both a user's and a developer's perspective. QMAP is publicly available as open-source at \url{https://github.com/cda-tum/qmap}.
\end{abstract}
\maketitle

\section{Introduction}\label{sec:introduction}
Quantum computing~\cite{nielsenQuantumComputationQuantum2010} employs quantum-mechanical principles to process data.
Unlike classical computing, which uses binary digits (bits) that can only be 0 or 1, quantum computing uses quantum bits. This allows for states in superposition or states that are entangled. 
Because of these quantum-mechanical properties, quantum computers can solve some computational problems exponentially faster than classical computers. 
One of the most notable tasks is determining the prime factors of a large composite number~\cite{shorPolynomialtimeAlgorithmsPrime1997}.
Because many modern encryption systems are based on this problem, resolving it quickly would have far-reaching consequences for code-breaking and cybersecurity.
Furthermore, quantum computing has the potential to transform fields such as machine learning~\cite{biamonteQuantumMachineLearning2017}, optimization~\cite{harwoodFormulatingSolvingRouting2021}, and quantum simulation~\cite{brownUsingQuantumComputers2010}.
The ability to perform specific tasks exponentially faster than traditional computers may also aid in drug discovery~\cite{caoPotentialQuantumComputing2018}, materials science~\cite{huangSimulatingElectronicStructure2022}, and artificial intelligence~\cite{dunjkoMachineLearningArtificial2018}.
However, quantum error correction, the development of quantum algorithms and software, as well as the physical realization of large-scale quantum computers, continue to be major challenges for the field. 
But quantum computing research is moving quickly, and quantum computer capabilities are growing continuously.

Commonly, quantum algorithms are described as quantum circuits, which are sequences of quantum operations (or gates) acting on qubits.
A quantum circuit compiler must be used to compile an algorithm before it can be executed on a real \emph{Quantum Processing Unit}~(QPU).
Typically, such a compiler begins by simplifying the original circuit description using device-independent optimizations~\cite{hattoriQuantumCircuitOptimization2018,sasanianReversibleQuantumCircuit2013,itokoOptimizationQuantumCircuit2020,maslovQuantumCircuitSimplification2008,schneiderSATEncodingOptimal2023,younisQuantumCircuitOptimization2022}.
These optimizations include techniques like canceling or rearranging gates and using high-level synthesis routines.
Following these preliminary optimizations, QPU-specific circuits are generated.
This is accomplished by translating all non-native gates into a set of native gates that can be directly executed on the device~\cite{vidalUniversalQuantumCircuit2004,maslovAdvantagesUsingRelative2016,millerElementaryQuantumGate2011,davisOptimalTopologyAware2020}.
Furthermore, some QPUs only feature limited connectivity between their qubits---necessitating a mapping step that maps the circuit's logical (or virtual) qubits to the device's (physical) qubits so that connectivity constraints are met throughout the circuit.

This work focuses on this quantum circuit mapping step. In fact, it is extremely difficult to ensure that all gates in a quantum circuit match the topology of a specified architecture while, at the same time, keeping the resulting costs (e.g., in terms of the number of additionally introduced gates, fidelity, etc.) low.
In practice, most quantum circuits cannot be easily mapped to existing quantum architectures.
As a result, additional gates (known as SWAP gates) are introduced during the mapping process to ensure that the circuit adheres to the architecture's connectivity constraints.
The quality of the resulting mapping can be assessed via a variety of metrics, such as the resulting circuit's (two-qubit) gate count, depth, or expected fidelity.
For the sake of clarity, we will focus on the number of additional gates in the following.
It is crucial for the performance of the resulting circuit to keep this value as low as possible, as it can frequently mean the difference between a circuit producing expected results and a circuit producing entirely random outputs.

This paper shows how to tackle the quantum circuit mapping problem efficiently using the open-source tool QMAP, which is part of the \emph{Munich Quantum Toolkit} (MQT)---a collection of quantum computing software tools whose main focus is on providing efficient, automated, and accessible methods for tackling challenging problems in the design of quantum computing applications.
To that end, we will provide a brief overview of various aspects, specifically
\begin{enumerate}
	\item a high-level description of the quantum circuit mapping problem and how the MQT QMAP tool helps solve it efficiently (covered in \autoref{sec:qmap}),
	\item the user's perspective on how QMAP can be used to efficiently map quantum circuits to quantum computing architectures (covered in \autoref{sec:users}), as well as,
	\item the developer's perspective on how to develop new or extend existing methods in the QMAP tool (covered in \autoref{sec:developers}).
\end{enumerate}

\section{Background}\label{sec:background}
Before delving into quantum circuit mapping and QMAP, this section provides a quick review of the fundamentals of quantum computing in general and current quantum computing architectures in particular.
 It should be noted that the scope of this summary paper does not allow for a comprehensive review of the broad field of quantum computation. We therefore direct anyone wishing to enter the field to a more in-depth treatment of the fundamentals as provided, for example, in~\cite{nielsenQuantumComputationQuantum2010}.

\subsection{Quantum Computing}\label{sec:qc}
In contrast to classical computing, which represents data using bits that can either be zero or one, quantum computing uses qubits, or quantum bits that can exist in a superposition of states, which means that they can be in a linear combination of the 0 and 1 states.
More specifically, the state $\ket{\psi}$ of a single qubit is described by two complex-valued amplitudes $\alpha_0$ and $\alpha_1$ such that
\[
\ket{\psi} = \alpha_0 \ket{0} + \alpha_1 \ket{1} \mbox{ with } \vert\alpha_0\vert^2 + \vert\alpha_1\vert^2 = 1.
\]
Measuring a qubit (the only way to extract information from a quantum computer), probabilistically collapses its state to one of the basis states $\ket{0}$ or $\ket{1}$ (with probability $\vert\alpha_0\vert^2$ or $\vert\alpha_1\vert^2$, respectively).

Just as in classical computing, multiple qubits are used to perform computations.
To this end, the state of an $n$-qubit system is described by $2^n$ complex-valued amplitudes $\alpha_i$; one for each possible bitstring of length $n$, i.e.,
\[
\ket{\psi} = \sum_{i\in\{0,1\}^n}\alpha_i \ket{i} \mbox{ with } \sum_{i\in\{0,1\}^n} \vert\alpha_i\vert^2 = 1.
\]
Again, $\vert\alpha_i\vert^2$ describes the probability of obtaining the basis state $\ket{i}$ when measuring the qubits.

The state of qubits is altered using quantum operations, also known as quantum gates.
Example of gates range from analogues of classical gates such as the X gate, which inverts the state of a qubit, to quantum-specific gates such as the Hadamard gate that can be used to put a qubit from a (classical) basis state into a (quantum) superposition.
These gates are used to construct quantum circuits, the quantum computing analogue of classical circuits.
Currently, quantum circuits are the predominant way to describe quantum computations.

\begin{figure}[t]
	\centering
	\includegraphics[width=0.99\linewidth]{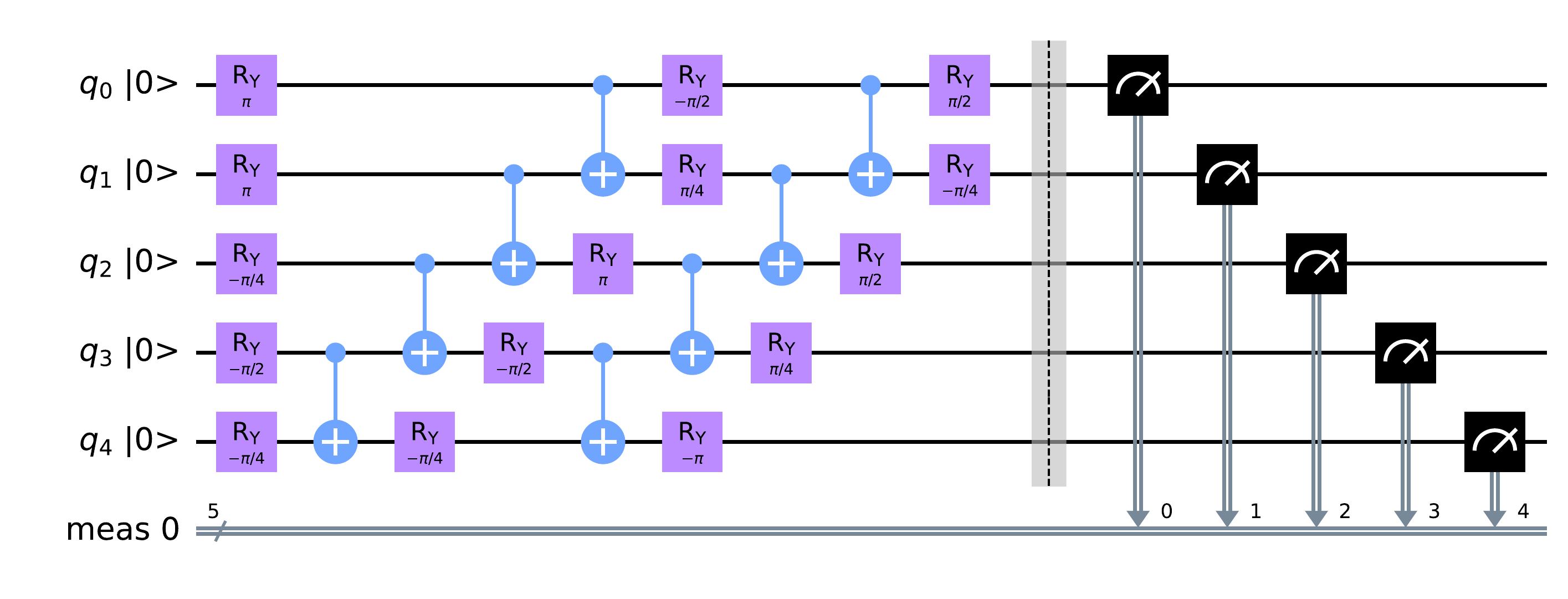}\vspace*{-3mm}
	\caption{Five-qubit VQE quantum circuit. The circuit consists of chains of two-qubit CNOT gates, interleaved with single-qubit rotation gates.}
	\label{fig:circ}
\end{figure}

\begin{example}\label{ex:circ}
	\autoref{fig:circ} shows a quantum circuit acting on five qubits.
	To this end, qubits are shown as horizontal lines and gates are denoted by boxes and symbols placed on these lines.
	Gates are applied in sequence from left to right.
	The circuit at hand consists of \mbox{single-qubit} rotation gates (indicated by purple boxes) and two-qubit CNOT gates (indicated in blue).
	It represents an instance of a popular near-term quantum algorithm (called \emph{Variational Quantum Eigensolver}, or VQE~\cite{cerezoVariationalQuantumAlgorithms2020}) applied to a small optimization problem.
\end{example}

\subsection{Quantum Computing Architectures}\label{sec:archs}

Several technologies are currently being utilized or developed for quantum computing.
Some of the most notable include superconducting qubits~\cite{krantzQuantumEngineerGuide2019}, trapped ions~\cite{haffnerQuantumComputingTrapped2008}, neutral atoms~\cite{henrietQuantumComputingNeutral2020}, and photonic qubits~\cite{obrienPhotonicQuantumTechnologies2009}. 
Each of these technologies has distinct advantages and disadvantages.
Superconducting qubits, for example, are very easy to control and manipulate, but they have to operate at extremely low temperatures.
Trapped ions, on the other hand, can operate at much higher temperatures and have a long coherence time, but they are difficult to scale to a large number of qubits.
It is crucial to note that most of the technologies are still in the research phase, and it is unclear which technology will prove to be most suitable for large-scale, general-purpose quantum computing.
The remainder of this paper focuses on quantum computers based on superconducting qubits, as the quantum circuit mapping problem is most relevant for them.

\begin{example}
	Major industrial players such as IBM, Rigetti, and Google as well as academic institution such as \emph{Oxford Quantum Computing}~(OQC) currently provide access to quantum computers based on superconducting qubits.
	At the end of 2022, IBM, for example,  provided devices ranging from $5$ qubits all the way up to $433$ qubits and they are planning for a system with more than $4000$ qubits by 2025 in their roadmap (\url{ibm.com/quantum/roadmap}).
\end{example}

Quantum computing architectures only offer a particular set of \emph{native gates}, typically consisting of some single-qubit operations and a two-qubit entangling operation---effectively forming a universal gate set with which any quantum computation can be realized.
This can be compared to a classical processor only supporting a particular \emph{Instruction Set Architecture} (ISA).
Just as a compiler is needed to convert high-level code to low-level instructions, high-level quantum circuits need to be decomposed to the native gate-set of the device they are supposed to be run on before being executable.

\begin{example}
	Again focusing on the quantum computers offered by IBM, they provide a native gate set that consists of three single-qubit gates ($X$, $\sqrt{X}$, and $R_z(\theta)$) as well as the two-qubit CNOT gate.
	Although seemingly small, this already constitutes a universal gate set for quantum computing, i.e., one out of which any conceivable quantum computation can be realized.
\end{example}

In addition, quantum computers, such as those based on superconducting qubits, only feature limited connectivity between their qubits, i.e., a two-qubit gate may not be applied to arbitrary qubits on the device. Usually, the ways in which qubits can interact with each other are described by a \emph{coupling graph} (or \emph{coupling map}). In general, this is a directed graph that describes which qubit pairs a certain two-qubit gate might be applied to. In most cases, the graph is actually undirected, meaning that the directionality of the gate between two qubits does not matter. To execute a circuit on one of these devices, the circuit has to respect the connectivity constraints imposed by the device’s coupling map.

\begin{figure}
	\centering
	\includegraphics[width=0.88\linewidth]{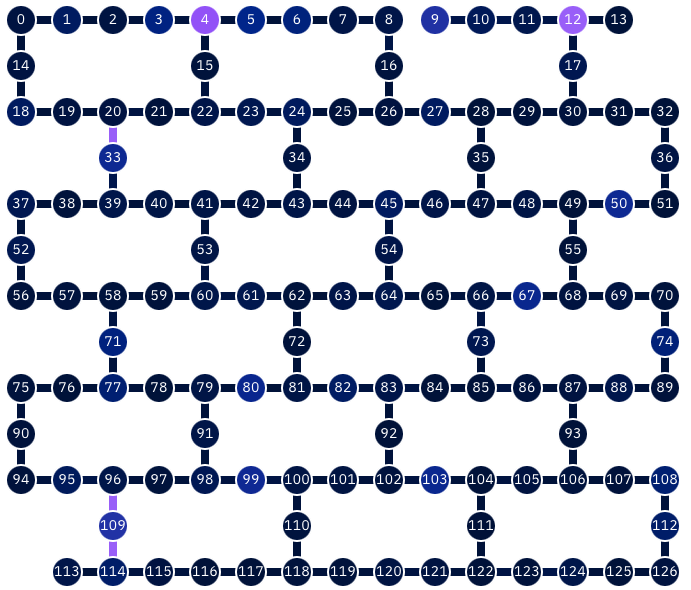}
	\caption{Coupling map of the $127$-qubit \emph{ibm\_washington} device}
	\label{fig:washington}
\end{figure}

\begin{example}
	\autoref{fig:washington} shows the coupling map of the $127$-qubit \emph{ibm\_washington} device.
	Like all recent IBM architectures, it is built from $12$-qubit tiles that are stitched together in a brick-like pattern.
	Note how sparsely connected the overall architecture is, with each qubit having at most three neighbours.
\end{example}

\section{Quantum Circuit Mapping \\and MQT QMAP}\label{sec:qmap}

As reviewed in the previous section, many quantum computing architectures limit the pairs of qubits that two-qubit operations can be applied to---commonly described by a device’s coupling map.
To execute a generic quantum circuit (with arbitrary interactions between its qubits) on such an architecture, the circuit needs to be mapped.
This involves \emph{qubit allocation}, where the circuit's (logical) qubits are assigned to the device's (physical) qubits in an initial layout, and \emph{routing}, where the original circuit is augmented with SWAP gates such that it adheres to the target device’s coupling map.

\begin{figure}[t]
	\centering
	\includegraphics[width=0.99\linewidth]{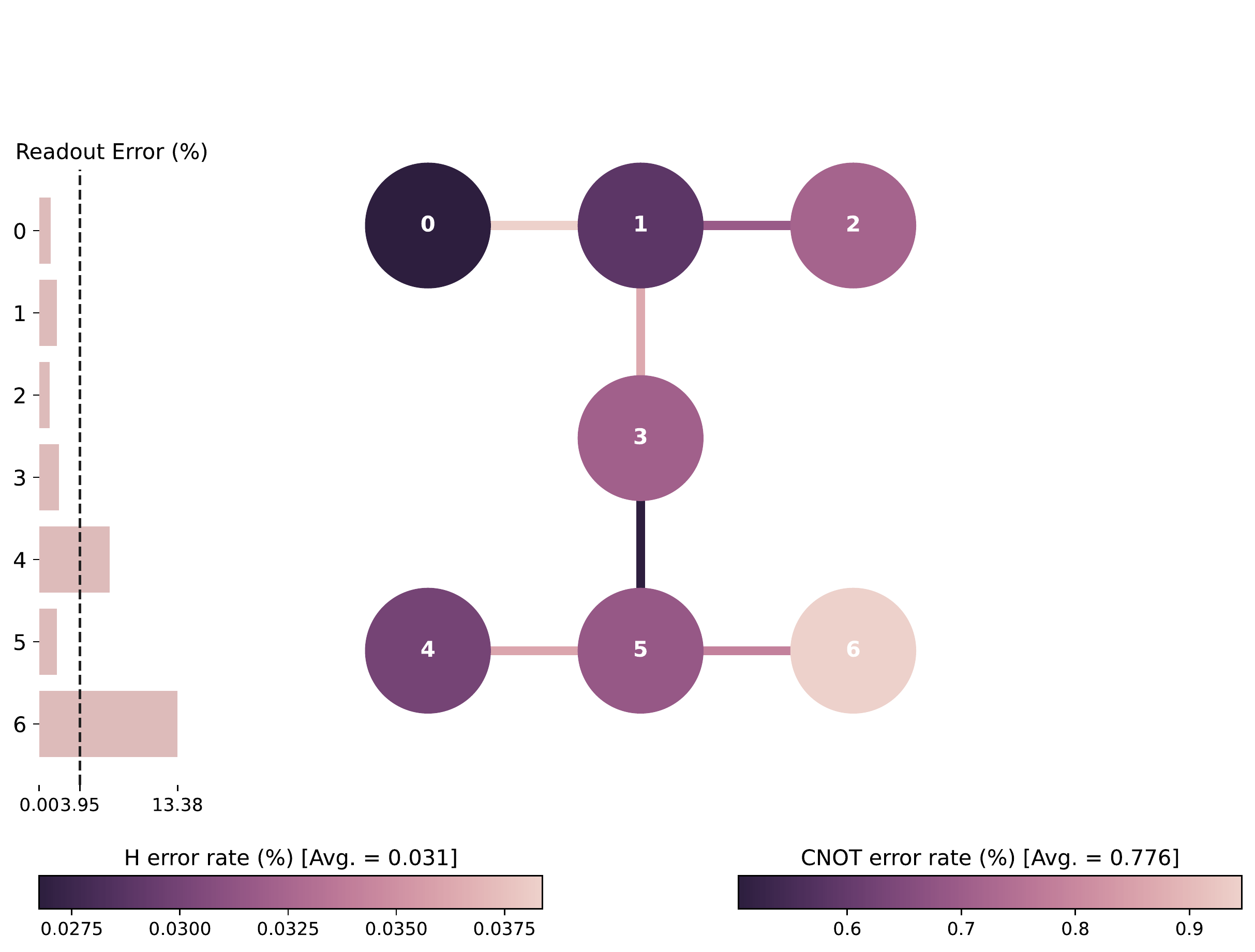}
	\caption{Seven-qubit \emph{ibm\_oslo} architecture with device calibration data. Circles represent the individual qubits. Connections between qubits indicate that a two-qubit gate may be applied to them. The color of the qubits (edges) indicates the error rate of single-qubit (two-qubit) operations applied to that qubit (edge)---with darker implying lower error rates.}
	\label{fig:arch}
\end{figure}

\begin{figure}[t]
	\centering
	\includegraphics[width=0.99\linewidth]{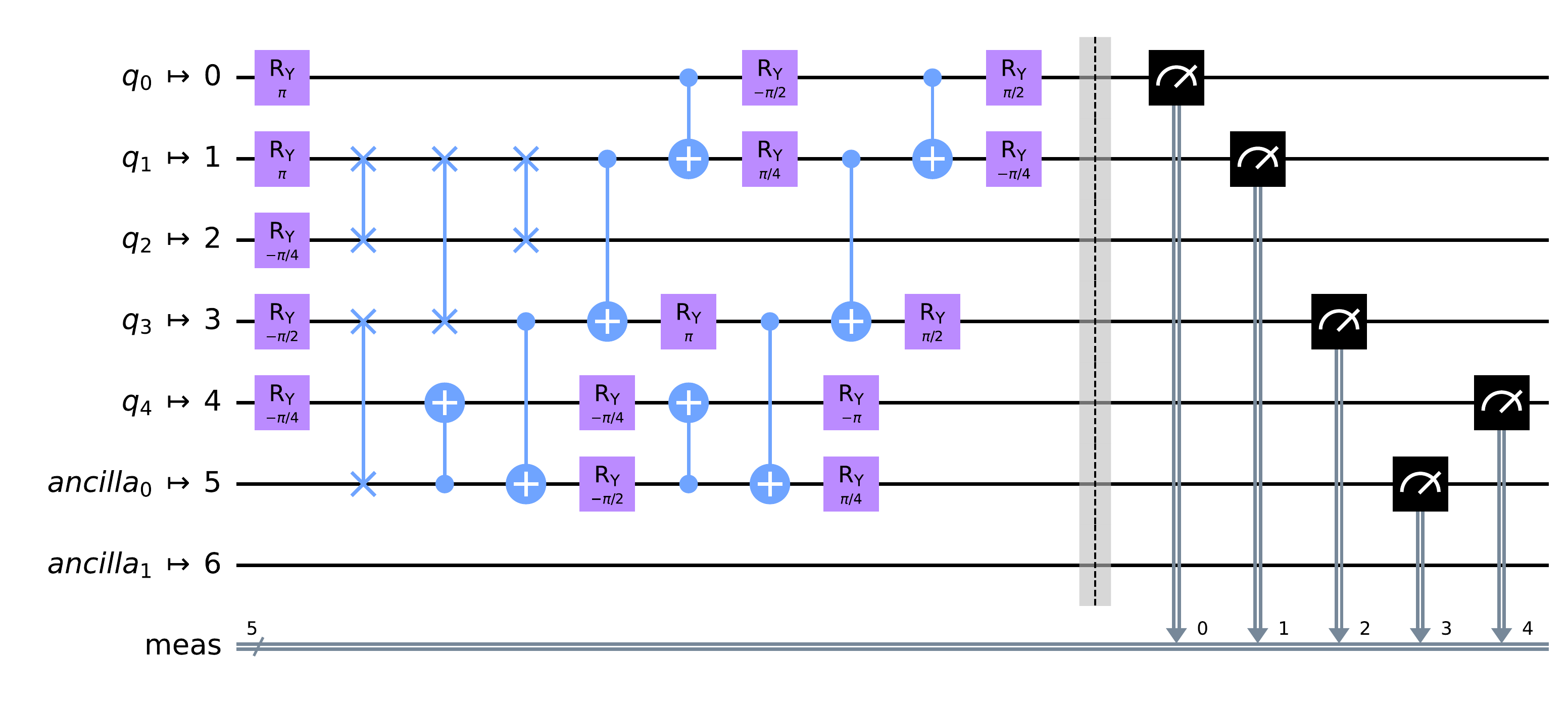}
	\caption{Naive mapping of the circuit from \autoref{fig:circ} to the \emph{ibm\_oslo} architecture shown in \autoref{fig:arch}. A total of four SWAP gates and an additional qubit are required to make the circuit executable.}
	\label{fig:naive}
\end{figure}

\begin{example}
	Consider the circuit from \autoref{ex:circ} (shown in \autoref{fig:circ}) and assume it shall be mapped to the seven-qubit \emph{ibm\_oslo} architecture whose coupling map (together with some calibration information) is shown in \autoref{fig:arch}. 
		This circuit cannot be run directly on the architecture since it contains gates that act on qubits not connected on the device's coupling map.
	Naively inserting SWAP gates that permute the logical-to-physical qubit mapping on the fly may yield a compiled circuit as shown in \autoref{fig:naive}.
	Over the course of the mapping, four SWAP gates have been introduced to satisfy the connectivity constraints of the device’s architecture.
	Furthermore, the circuit uses an additional qubit for realizing the intended functionality.
	Since every additional gate (and qubit) increases the probability of errors, this is a very costly overhead for such a small circuit.
\end{example}

Keeping the number of additionally introduced gates as small as possible during quantum circuit mapping is key for ensuring the successful execution of the quantum circuit.
Determining an optimal mapping for a quantum circuit is an NP-hard problem~\cite{boteaComplexityQuantumCircuit2018}.

At the time of writing, the open-source quantum circuit mapping tool QMAP, which is part of the \emph{Munich Quantum Toolkit}~(MQT), offers two dedicated techniques for tackling that problem:
\begin{enumerate}
	\item An exact mapping approach (based on \cite{willeMappingQuantumCircuits2019,burgholzerLimitingSearchSpace2022}) that guarantees (gate-optimal) solutions.
	\item A heuristic mapping approach (based on \cite{zulehnerEfficientMethodologyMapping2019,hillmichExploitingQuantumTeleportation2021,zulehnerCompilingSUQuantum2019}) that allows to determine efficient mapping solutions in a scalable fashion.
\end{enumerate}

The exact mapper implemented in QMAP maps quantum circuits using the minimal number of SWAP gates.
To this end, it encodes the mapping task as a MaxSAT problem and subsequently solves it using the SMT solver Z3~\cite{demouraZ3EfficientSMT2008}.
Due to the NP-hardness of the mapping task, this approach is only scalable up to roughly eight qubits and on the order of $1000$ gates in most scenarios.

\begin{figure}
	\centering
	\includegraphics[width=0.8\linewidth]{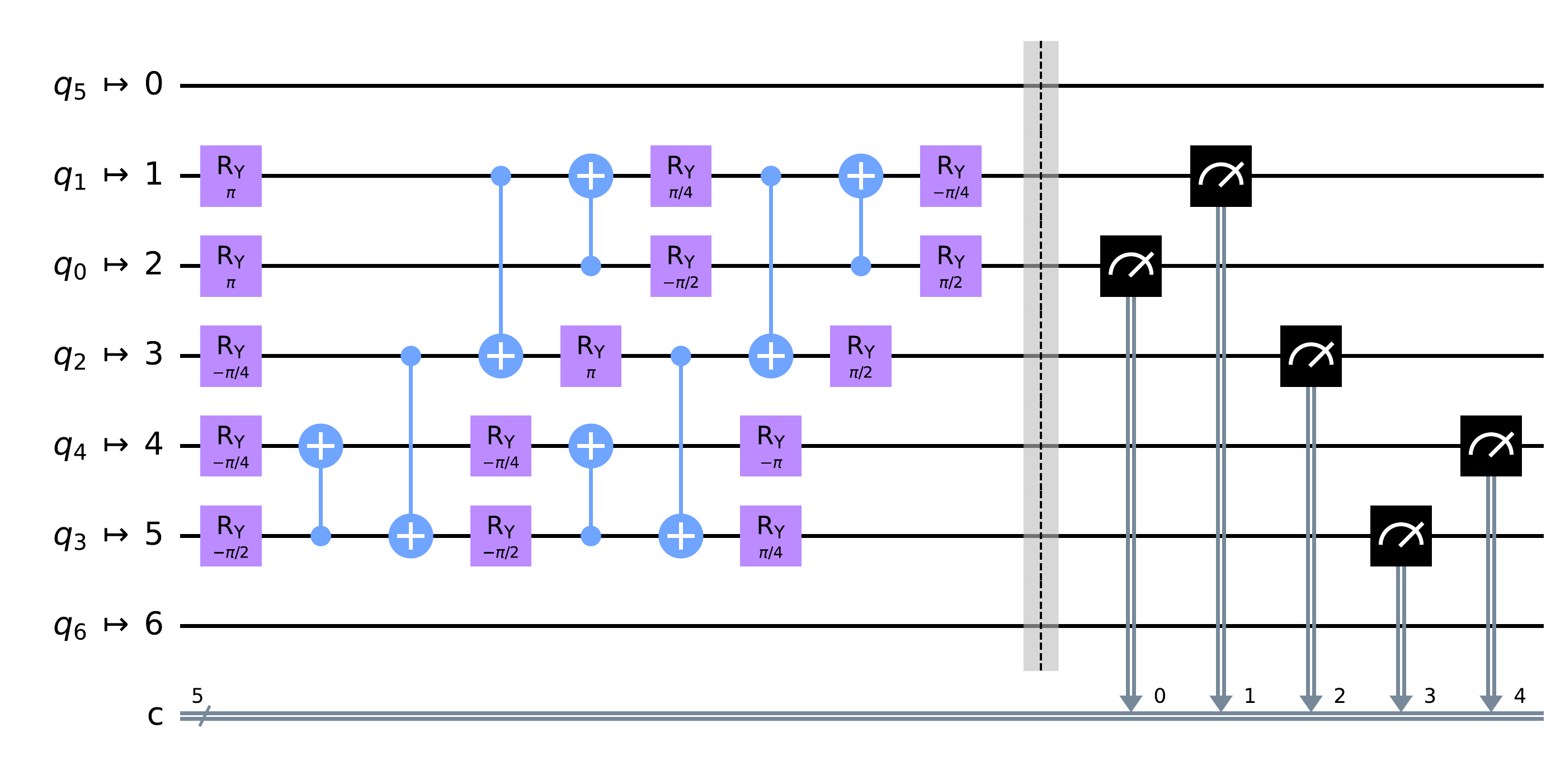}\vspace*{-3mm}
	\caption{Mapping of the circuit from \autoref{fig:circ} to the \emph{ibm\_oslo} architecture shown in \autoref{fig:arch} using the exact mapper of QMAP. No SWAP gates or additional qubits are required to make the circuit executable.}
	\label{fig:optimal}
\end{figure}

\begin{example}
	When used on the circuit shown in \autoref{fig:circ}, the exact mapper of QMAP yields the circuit shown in \autoref{fig:optimal}.
	In this case, not a single SWAP gate and no additional qubits are required to map the circuit.
	Consequently, the expected performance of the resulting circuit is significantly improved.
	This performance comes at the cost of a mapping runtime in the order of a couple of seconds.
\end{example}

The heuristic mapper implemented in QMAP uses $A^*$-search~\cite{hartFormalBasisHeuristic1968} to efficiently traverse the immense search space of the mapping problem.
It effectively trades optimality for runtime.
This allows to reliably determine suitable mappings for circuits with up to hundreds of qubits and hundreds of thousands of gates.

\begin{figure}
	\centering
	\includegraphics[width=0.9\linewidth]{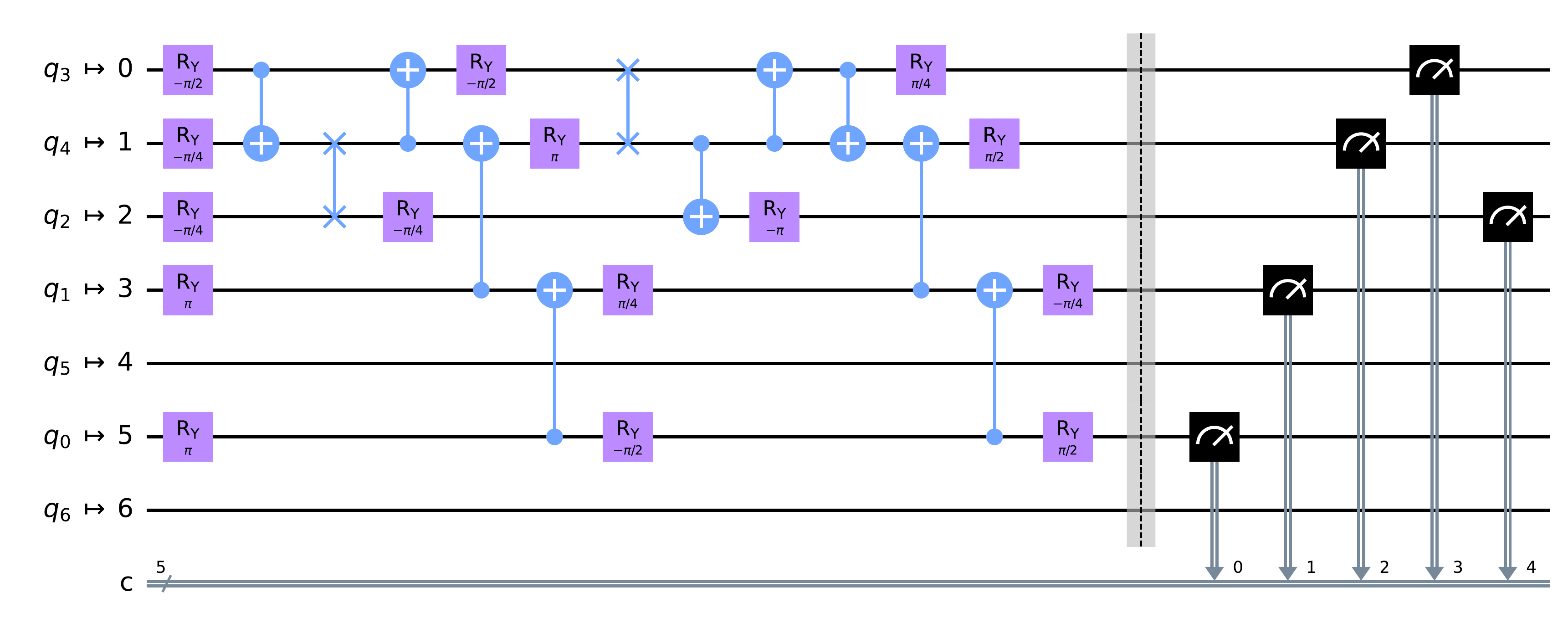}
	\caption{Mapping of the circuit from \autoref{fig:circ} to the \emph{ibm\_oslo} architecture shown in \autoref{fig:arch} using the heuristic mapper of QMAP. The mapping uses two SWAP gates and one additional CNOT gate.}
	\label{fig:heuristic}
\end{figure}

\begin{example}
	When used on the circuit shown in \autoref{fig:circ}, the heuristic mapper of QMAP yields the circuit shown in \autoref{fig:heuristic}.
	While this solution is not optimal anymore it still only requires two SWAP gates and an additional CNOT gate as opposed to the four SWAP gates in the naive mapping.
	Furthermore, even for such a small example the heuristic mapper is orders of magnitudes faster than the exact mapper---requiring less than a millisecond.
\end{example}

\section{User's Perspective: \\Working with QMAP}\label{sec:users}

Anyone may quickly and effectively map their quantum circuits to quantum computing systems with QMAP.
The installation process and initial use of the MQT QMAP tool are briefly described in this section.
By doing this, we offer a view into QMAP from the user's perspective.

QMAP is mainly developed as a C++ library but, in order to make the tool as accessible and compatible as possible, comes with an easy-to-use Python interface.
Installing the tool is as easy as
\begin{lstlisting}[language=bash]
	$ python3 -m venv .venv
	$ source venv/bin/activate
	(venv) $ pip install -U pip setuptools wheel
	(venv) $ pip install mqt.qmap
\end{lstlisting}
In most practical cases (under 64-bit Linux, MacOS incl. Apple Silicon, and Windows), this requires no compilation and merely downloads and installs a platform-specific pre-built wheel.

The architecture to which a particular circuit should be mapped can be specified in a number of different ways.
The simplest option is to just use one of IBM's Qiskit's~\cite{treinishQiskitQiskit2022} backends, but (arbitrary) architectures can also be created directly from a coupling map.

\begin{example}
The \emph{ibm\_oslo} device (shown in \autoref{fig:arch}) considered previously can be instantiated via
\begin{lstlisting}[language=Python]
	from qiskit.providers.fake_provider import FakeOslo

	arch = FakeOslo()
\end{lstlisting}
The same architecture could have also been provided without relying on Qiskit via
\begin{lstlisting}[language=Python]
	from mqt import qmap

	arch = qmap.Architecture(7,{
		(0, 1),	(1, 0), 
		(1, 2), (2, 1),	
		(1, 3), (3, 1),	
		(2, 3), (3, 2), 
		(3, 5), (5, 3),	
		(4, 5), (5, 4),	
		(5, 6), (6, 5)}
	)
\end{lstlisting}
\end{example}

With that, it is possible to define any desired quantum circuit and map it to the architecture.
Once more, using Qiskit in Python is the easiest method.
Alternative file formats, such OpenQASM~\cite{crossOpenQASMBroaderDeeper2021}, also can be used to define the circuit.

\begin{example}
	The circuit shown in \autoref{fig:circ} originates from the MQT Bench library~\cite{quetschlichMQTBenchBenchmarking2022} which can be installed via
\begin{lstlisting}[language=bash]
	(venv) $ pip install mqt.bench
\end{lstlisting}
	Then, the circuit can be obtained as a Qiskit \texttt{QuantumCircuit} via
\begin{lstlisting}[language=Python]
	from mqt import bench

	qc = bench.get_benchmark(benchmark_name='vqe', circuit_size=5, level='indep')
\end{lstlisting}
Note that the particular parameters obtained for the rotation gates may vary with each invocation.
\end{example}

Once the architecture \texttt{arch} to map to and the quantum circuit \texttt{qc} to map have been defined, actually performing the mapping itself using QMAP only requires a single line of Python 
\begin{lstlisting}[language=Python]
	qc_m, res = qmap.compile(qc, arch, method="exact")
\end{lstlisting}
for the exact mapper, or
\begin{lstlisting}[language=Python]
	qc_m, res = qmap.compile(qc, arch, method="heuristic")
\end{lstlisting}
for the heuristic mapper.
Both return the mapped circuit as an IBM Qiskit \texttt{QuantumCircuit} object and a collection of results.
The results, for example, allow one to conveniently print the required number of SWAP operations and the runtime of the mapping routine via
\begin{lstlisting}[language=Python]
	print("Additional SWAPs: %d" % res.output.swaps)
	print("Runtime:          %f" % res.time)
\end{lstlisting}

Naturally, the aforementioned walk-through only offered a brief overview of QMAP's capabilities.
The \texttt{compile} method has many optional arguments that can be used to configure the individual mappers.
However, QMAP has comprehensive documentation that can be accessed at \url{https://mqtqmap.readthedocs.io} and offers a more extensive overview of all available options and techniques.

\vspace*{-3mm}
\section{Developer's Perspective: \\Improving QMAP}\label{sec:developers}\vspace*{-1mm}

QMAP is a powerful tool, but it could be improved in many ways.
Many aspects of QMAP were designed from a purely academic perspective with a rather narrow focus on very specific topics.
This section begins with a quick overview of how to set up a development environment for QMAP, followed by a discussion of potential enhancement areas.
More specifically, we show
\begin{itemize}
	\item how the heuristic mapper could be made better by defining more complex initial layout algorithms as well as allowing for a wider variety of gates, and
	\item how the exact mapper's SAT encoding and its construction could be adapted to improve the mapper's overall performance.
\end{itemize}
This provides a brief look at QMAP from the developer's perspective.

\vspace*{-3mm}
\subsection{Setting up a Development Environment}\label{sec:devel}\vspace{-1mm}

QMAP is mainly developed as a C++ library that builds on the MQT \emph{Quantum Functionality Representation} (QFR) library (\url{https://github.com/cda-tum/qfr}).
In order to make the tool as accessible and compatible as possible, it comes with easy-to-use Python bindings.
Consequently, developing features for the QMAP library mostly entails extending the core C++ library and, then, exposing the newly-added functionality to Python.

To get started, the project needs to be cloned from GitHub via
\begin{lstlisting}[language=bash]
	git clone --recursive https://github.com/cda-tum/qmap
\end{lstlisting}
Building the project requires a C++ compiler supporting C++17 and CMake with a minimum version of 3.19.
In addition, the SMT solver Z3~\cite{demouraZ3EfficientSMT2008} needs to be installed with a minimum version of 4.8.15 to build the exact mapper (see \url{https://github.com/Z3Prover/z3} for further details).
Just as any other CMake-based project, building the C++ library is a two-step procedure
\begin{lstlisting}[language=bash]
	[.../qmap] $ cmake -S . -B build -DBUILD_QMAP_TESTS=ON
	[.../qmap] $ cmake --build build
\end{lstlisting}

To locally build and install the Python package, first create a virtual environment as described in the previous section and, then, call
\begin{lstlisting}[language=bash]
	[.../qmap] (venv) $ pip install -e ".[dev]"
\end{lstlisting}

The library provides many convenience features, such as unit tests to test the library's functionality, linting and code formatting tools to ensure proper quality of the code, and a comprehensive setup of CI workflows on GitHub that ensure the project builds and runs on all major operating systems. For further information, visit \url{mqtqmap.readthedocs.io/en/latest/DevelopmentGuide.html}.

Having the development environment set up, QMAP can easily be extended and, by this, enhanced. In the following, examples for both, the heuristic mapper and the exact mapper are provided. 

\subsection{Improving the Heuristic Mapper}\label{sec:heur}

One direction how the heuristic mapper can be improved involves the initial mapping.
As most quantum circuit mappers, the heuristic QMAP mapper splits the mapping tasks into (1)~determining an initial layout for the placement of the logical qubits on the device's physical qubits and (2) performing the actual routing based on the chosen initial layout.
At the time of writing, QMAP supports three different kinds of initial layouts:
\begin{enumerate}
	\item A trivial \emph{identity} layout, where logical qubit $q_i$ is assigned to physical qubit $Q_i$ for all $i$ from $0$ to $n-1$.
	\item A \emph{static} layout, which considers qubits that share a gate in the first layer of the circuit and maps those to any free connected qubit pair in the architecture while mapping the remaining qubits by order of index.
	\item A \emph{dynamic} layout, where qubits are greedily assigned \mbox{on-demand} when encountering them during the mapping.
\end{enumerate}
As evaluations in the corresponding literature confirm~\cite{zulehnerEfficientMethodologyMapping2019,zulehnerCompilingSUQuantum2019,hillmichExploitingQuantumTeleportation2021}, these initial layout techniques in combination with the $A^*$ search already allow for rather efficient mapping results.
At the same time, it is well known that the initial mapping has a significant impact on the routing performance that can be achieved; see, e.g.,~\cite{palerInfluenceInitialQubit2019}.
Consequently, more sophisticated initial mappings may lead to even better overall results---offering room for improvement.
Many tools for mapping quantum circuits today (such as~\cite{sivarajahKetRetargetableCompiler2021}) determine an initial mapping by focusing on matching a graph of the circuit's gate interactions to the coupling map of the target architecture.
This can be supplemented with device calibration information to account for differences in qubit and gate quality.
Implementing such tactics could be a straightforward means of improving QMAP's efficiency even further.

Another direction could involve the supported gate sets.
For historical reasons, QMAP assumes that the only two-qubit gates in the circuit are CNOT gates.
This was a reasonable assumption at a time when only IBM provided public access to their devices and Qiskit was the only major open-source quantum SDK.
In the meantime, however, many further (and different) hardware providers and quantum computers have emerged---each with their own native gate set. 
As an example, Rigetti quantum computers expose the CZ gate rather than the CNOT gate.
As a result, adapting QMAP to newer types of devices and/or to support more diverse gate sets is another straightforward means of improving QMAP.

\subsection{Improving the Exact Mapper}\label{sec:exact}

Also the exact mapper offers room for improvements.
As previously discussed, the exact mapper encodes the mapping task as a MaxSAT problem, allowing it to deliver results with the fewest possible SWAP gates. The quality of the results comes at a cost, however, because the search space of the mapping problem scales as $N!*k$, where $N$ is the number of qubits in the architecture and $k$ is the number of (two-qubit) gates in the circuit. As a result, the technique is currently limited to architectures with fewer than ten qubits and circuits with a few hundred to a thousand gates.

A number of efforts have been made to prune the search space in the quantum circuit mapping problem. It has been demonstrated in \cite{burgholzerLimitingSearchSpace2022} that it is sufficient to consider just enough permutations in front of each layer in the mapping for any two qubits to be placed adjacently. This drastically reduces the number of permutations from $N!$. Nonetheless, the concepts described in \cite{burgholzerLimitingSearchSpace2022} are implemented in a rather straightforward manner. Further improvements could be made in this area by streamlining the computation of the permutations that must be considered, for example, by developing a specialized Cayley graph datastructure and corresponding manipulation algorithms.

In addition, attempts have been made to limit the number of qubits that must be considered during the mapping.
Without such restrictions, it would be impossible to map a 3-qubit circuit to a 100-qubit architecture, for instance.
The most radical approach is given in~\cite{willeMappingQuantumCircuits2019}, which considers all viable subarchitectures containing as many qubits as the circuit to be mapped rather than the entire architecture.
While this reduces the number of qubits to the greatest extent possible (a circuit can never be mapped to an architecture with fewer qubits than the circuit itself), it is not without its own downsides.
In the first place, the existing implementation is relatively simple, as it does not account for subarchitecture isomorphism, i.e., the possibility that many subarchitectures are structurally identical.
Instead, it simply explores every possible combination using brute force.
Furthermore, it has been demonstrated in \cite{pehamOptimalSubarchitecturesQuantum2022} that the optimality of the resulting circuit cannot be guaranteed because subarchitectures with more qubits may permit a better mapping.
Fortunately, \cite{pehamOptimalSubarchitecturesQuantum2022} proposes methods for determining optimal or near-optimal sets of subarchitectures to be considered during mapping.
These methods have not yet been incorporated directly into the exact mapper.

\section{Conclusions}

This summary paper described the MQT QMAP tool, which allows researchers and developers to effectively map quantum circuits to real quantum computing architectures. We briefly reviewed the quantum circuit mapping problem and demonstrated how QMAP can be used to solve this problem efficiently and in an automated fashion. Furthermore, we gave a rough overview of how to develop for QMAP and offered potential avenues for future improvement. In doing so, we hope to pique more interest in this intriguing and rapidly growing area of research. In that case, we are referring to the aforementioned references for a more in-depth discussion of the relevant methods and challenges.

\vspace{4mm}
\subsection*{Acknowledgements}
We thank everyone that contributed to the QMAP tool.
Special thanks to Stefan Hillmich, Tom Peham, Sarah Schneider, and Alwin Zulehner for their specific contributions to the work presented here.

This work received funding from the European Research Council (ERC) under the European Union’s Horizon 2020 research and innovation program (grant agreement No. $101001318$), was part of the Munich Quantum Valley, which is supported by the Bavarian state government with funds from the Hightech Agenda Bayern Plus, and has been supported by the BMWK on the basis of a decision by the German Bundestag through project QuaST.

\printbibliography

@STRING{tcad	= {{IEEE} Trans. on {CAD} of Integrated Circuits and Systems} }

@STRING{siam	= {SIAM Jour. of Comp.} }

@STRING{tqe     = {IEEE Transactions on Quantum Engineering} }

@STRING{dac	= {Design Automation Conf.} }

@STRING{aspdac	= {Asia and South Pacific Design Automation Conf.} }

@STRING{date	= {Design, Automation and Test in Europe} }

@STRING{control	= {Control Conference} }

@STRING{rc_conf	= {Int'l Conf. of Reversible Computation} }

@STRING{ismvl	= {Int'l Symp. on {M}ulti-{V}alued {L}ogic} }

@STRING{sat     = {Conference on Theory and Applications of Satisfiability Testing}}

@article{biamonteQuantumMachineLearning2017,
  title = {Quantum machine learning},
  author = {Biamonte, Jacob and Wittek, Peter and Pancotti, Nicola and Rebentrost, Patrick and Wiebe, Nathan and Lloyd, Seth},
  date = {2017-09},
  journaltitle = {Nature},
  volume = {549},
  number = {7671},
  eprint = {1611.09347},
  eprinttype = {arxiv},
  pages = {195--202},
  archiveprefix = {arXiv}
}

@inproceedings{boteaComplexityQuantumCircuit2018,
  title = {On the complexity of quantum circuit compilation},
  booktitle = {Intl {{Symp Comb}}. {{Search}}},
  author = {Botea, A. and Kishimoto, A. and Marinescu, Radu},
  date = {2018},
  eventtitle = {Int'l {{Symp}}. on {{Combinatorial Search}}}
}

@article{brownUsingQuantumComputers2010,
  title = {Using quantum computers for quantum simulation},
  author = {Brown, Katherine L. and Munro, William J. and Kendon, Vivien M.},
  date = {2010-11-15},
  journaltitle = {Entropy},
  volume = {12},
  number = {11},
  pages = {2268--2307},
  doi = {10.3390/e12112268},
  langid = {english}
}

@inproceedings{burgholzerLimitingSearchSpace2022,
  title = {Limiting the search space in optimal quantum circuit mapping},
  booktitle = aspdac,
  author = {Burgholzer, Lukas and Schneider, Sarah and Wille, Robert},
  date = {2022-01-17},
  eventtitle = {aspdac}
}

@article{caoPotentialQuantumComputing2018,
  title = {Potential of quantum computing for drug discovery},
  author = {Cao, Y. and Romero, J. and Aspuru-Guzik, A.},
  date = {2018-11-01},
  journaltitle = {IBM J. Res. \& Dev.},
  volume = {62},
  number = {6},
  pages = {6:1-6:20},
  doi = {10.1147/JRD.2018.2888987}
}

@misc{cerezoVariationalQuantumAlgorithms2020,
  title = {Variational quantum algorithms},
  author = {Cerezo, M. and Arrasmith, Andrew and Babbush, Ryan and Benjamin, Simon C. and Endo, Suguru and Fujii, Keisuke and McClean, Jarrod R. and Mitarai, Kosuke and Yuan, Xiao and Cincio, Lukasz and Coles, Patrick J.},
  date = {2020-12-16},
  eprint = {2012.09265},
  eprinttype = {arxiv},
  archiveprefix = {arXiv}
}

@misc{crossOpenQASMBroaderDeeper2021,
  title = {{{OpenQASM}} 3: {{A}} broader and deeper quantum assembly language},
  author = {Cross, Andrew W. and Javadi-Abhari, Ali and Alexander, Thomas and family=Beaudrap, given=Niel, prefix=de, useprefix=true and Bishop, Lev S. and Heidel, Steven and Ryan, Colm A. and Smolin, John and Gambetta, Jay M. and Johnson, Blake R.},
  date = {2021},
  eprint = {2104.14722},
  eprinttype = {arxiv},
  archiveprefix = {arXiv}
}

@inproceedings{davisOptimalTopologyAware2020,
  title = {Towards {{Optimal Topology Aware Quantum Circuit Synthesis}}},
  booktitle = {qce},
  author = {Davis, Marc G. and Smith, Ethan and Tudor, Ana and Sen, Koushik and Siddiqi, Irfan and Iancu, Costin},
  date = {2020-10},
  pages = {223--234},
  publisher = {{IEEE}},
  location = {{Denver, CO, USA}},
  doi = {10.1109/QCE49297.2020.00036},
  eventtitle = {2020 {{IEEE International Conference}} on {{Quantum Computing}} and {{Engineering}} ({{QCE}})}
}

@inproceedings{demouraZ3EfficientSMT2008,
  title = {Z3: {{An}} efficient {{SMT}} solver},
  shorttitle = {Z3},
  booktitle = {Tools {{Algorithms Constr}}. {{Anal}}. {{Syst}}.},
  author = {family=Moura, given=Leonardo, prefix=de, useprefix=true and Bjørner, Nikolaj},
  editor = {Ramakrishnan, C. R. and Rehof, Jakob},
  date = {2008},
  pages = {337--340},
  publisher = {{Springer}}
}

@article{dunjkoMachineLearningArtificial2018,
  title = {Machine learning \& artificial intelligence in the quantum domain: a review of recent progress},
  shorttitle = {Machine learning \& artificial intelligence in the quantum domain},
  author = {Dunjko, Vedran and Briegel, Hans J},
  date = {2018-07-01},
  journaltitle = {Rep. Prog. Phys.},
  volume = {81},
  number = {7},
  pages = {074001},
  doi = {10.1088/1361-6633/aab406}
}

@article{haffnerQuantumComputingTrapped2008,
  title = {Quantum computing with trapped ions},
  author = {Haffner, H and Roos, C and Blatt, R},
  date = {2008-12},
  journaltitle = {Physics Reports},
  volume = {469},
  number = {4},
  pages = {155--203},
  doi = {10.1016/j.physrep.2008.09.003},
  langid = {english}
}

@article{hartFormalBasisHeuristic1968,
  title = {A {{Formal Basis}} for the {{Heuristic Determination}} of {{Minimum Cost Paths}}},
  author = {Hart, Peter and Nilsson, Nils and Raphael, Bertram},
  date = {1968},
  journaltitle = {IEEE Trans. Syst. Sci. Cyber.},
  volume = {4},
  number = {2},
  pages = {100--107},
  doi = {10.1109/TSSC.1968.300136}
}

@article{harwoodFormulatingSolvingRouting2021,
  title = {Formulating and solving routing problems on quantum computers},
  author = {Harwood, Stuart and Gambella, Claudio and Trenev, Dimitar and Simonetto, Andrea and Bernal Neira, David and Greenberg, Donny},
  date = {2021},
  journaltitle = tqe,
  volume = {2},
  pages = {1--17},
  doi = {10.1109/TQE.2021.3049230}
}

@inproceedings{hattoriQuantumCircuitOptimization2018,
  title = {Quantum circuit optimization by changing the gate order for {{2D}} nearest neighbor architectures},
  booktitle = rc_conf,
  author = {Hattori, Wakaki and Yamashita, Shigeru},
  editor = {Kari, Jarkko and Ulidowski, Irek},
  date = {2018},
  volume = {11106},
  pages = {228--243},
  doi = {10.1007/978-3-319-99498-7_16},
  eventtitle = {rc\_conf}
}

@article{henrietQuantumComputingNeutral2020,
  title = {Quantum computing with neutral atoms},
  author = {Henriet, Loïc and Beguin, Lucas and Signoles, Adrien and Lahaye, Thierry and Browaeys, Antoine and Reymond, Georges-Olivier and Jurczak, Christophe},
  date = {2020-09-21},
  journaltitle = {Quantum},
  volume = {4},
  pages = {327},
  doi = {10.22331/q-2020-09-21-327},
  langid = {english}
}

@inproceedings{hillmichExploitingQuantumTeleportation2021,
  title = {Exploiting {{Quantum Teleportation}} in {{Quantum Circuit Mapping}}},
  booktitle = date,
  author = {Hillmich, Stefan and Zulehner, Alwin and Wille, Robert},
  date = {2021-01-18},
  pages = {792--797},
  publisher = {{ACM}},
  location = {{Tokyo Japan}},
  doi = {10.1145/3394885.3431604},
  eventtitle = {date},
  langid = {english}
}

@article{huangSimulatingElectronicStructure2022,
  title = {Simulating the {{Electronic Structure}} of {{Spin Defects}} on {{Quantum Computers}}},
  author = {Huang, Benchen and Govoni, Marco and Galli, Giulia},
  date = {2022-03-10},
  journaltitle = {PRX Quantum},
  volume = {3},
  number = {1},
  pages = {010339},
  doi = {10.1103/PRXQuantum.3.010339},
  langid = {english}
}

@article{itokoOptimizationQuantumCircuit2020,
  title = {Optimization of quantum circuit mapping using gate transformation and commutation},
  author = {Itoko, Toshinari and Raymond, Rudy and Imamichi, Takashi and Matsuo, Atsushi},
  date = {2020-01-01},
  journaltitle = {Integration},
  volume = {70},
  pages = {43--50},
  doi = {10.1016/j.vlsi.2019.10.004}
}

@article{krantzQuantumEngineerGuide2019,
  title = {A quantum engineer's guide to superconducting qubits},
  author = {Krantz, P. and Kjaergaard, M. and Yan, F. and Orlando, T. P. and Gustavsson, S. and Oliver, W. D.},
  date = {2019-06},
  journaltitle = {Applied Physics Reviews},
  volume = {6},
  number = {2},
  pages = {021318},
  doi = {10.1063/1.5089550},
  langid = {english}
}

@article{maslovAdvantagesUsingRelative2016,
  title = {On the advantages of using relative phase {{Toffolis}} with an application to multiple control {{Toffoli}} optimization},
  author = {Maslov, Dmitri},
  date = {2016-02-10},
  journaltitle = {Phys. Rev. A},
  volume = {93},
  number = {2},
  pages = {022311},
  doi = {10.1103/PhysRevA.93.022311}
}

@article{maslovQuantumCircuitSimplification2008,
  title = {Quantum circuit simplification and level compaction},
  author = {Maslov, D. and Dueck, G.W. and Miller, D.M. and Negrevergne, C.},
  date = {2008-03},
  journaltitle = tcad,
  volume = {27},
  number = {3},
  pages = {436--444},
  doi = {10.1109/TCAD.2007.911334}
}

@inproceedings{millerElementaryQuantumGate2011,
  title = {Elementary quantum gate realizations for multiple-control {{Toffoli}} gates},
  booktitle = ismvl,
  author = {Miller, D. Michael and Wille, Robert and Sasanian, Zahra},
  date = {2011},
  doi = {10.1109/ISMVL.2011.54},
  eventtitle = {ismvl}
}

@book{nielsenQuantumComputationQuantum2010,
  title = {Quantum {{Computation}} and {{Quantum Information}}},
  shorttitle = {Quantum {{Computation}} and {{Quantum Information}}},
  author = {Nielsen, Michael A. and Chuang, Isaac L.},
  date = {2010},
  publisher = {{Cambridge University Press}}
}

@article{obrienPhotonicQuantumTechnologies2009,
  title = {Photonic quantum technologies},
  author = {O'Brien, Jeremy L. and Furusawa, Akira and Vučković, Jelena},
  date = {2009-12},
  journaltitle = {Nature Photon},
  volume = {3},
  number = {12},
  pages = {687--695},
  doi = {10.1038/nphoton.2009.229},
  langid = {english}
}

@inproceedings{palerInfluenceInitialQubit2019,
  title = {On the {{Influence}} of {{Initial Qubit Placement During NISQ Circuit Compilation}}},
  booktitle = {Quantum {{Technol}}. {{Optim}}. {{Probl}}.},
  author = {Paler, Alexandru},
  date = {2019},
  series = {Lecture {{Notes}} in {{Computer Science}}},
  volume = {11413},
  pages = {207--217},
  doi = {10.1007/978-3-030-14082-3_18},
  eventtitle = {Quantum {{Technology}} and {{Optimization Problems}}},
  langid = {english}
}

@misc{pehamOptimalSubarchitecturesQuantum2022,
  title = {On {{Optimal Subarchitectures}} for {{Quantum Circuit Mapping}}},
  author = {Peham, Tom and Burgholzer, Lukas and Wille, Robert},
  date = {2022-10-17},
  number = {arXiv:2210.09321},
  eprint = {2210.09321},
  eprinttype = {arxiv},
  primaryclass = {quant-ph},
  publisher = {{arXiv}},
  url = {http://arxiv.org/abs/2210.09321},
  urldate = {2022-12-13},
  archiveprefix = {arXiv}
}

@misc{quetschlichMQTBenchBenchmarking2022,
  title = {{{MQT Bench}}: {{Benchmarking}} software and design automation tools for quantum computing},
  shorttitle = {{{MQT Bench}}},
  author = {Quetschlich, Nils and Burgholzer, Lukas and Wille, Robert},
  date = {2022-04-28},
  eprint = {2204.13719},
  eprinttype = {arxiv},
  archiveprefix = {arXiv}
}

@inproceedings{sasanianReversibleQuantumCircuit2013,
  title = {Reversible and quantum circuit optimization: {{A}} functional approach},
  shorttitle = {Reversible and quantum circuit optimization},
  booktitle = rc_conf,
  author = {Sasanian, Zahra and Miller, D. Michael},
  editor = {Glück, Robert and Yokoyama, Tetsuo},
  date = {2013},
  volume = {7581},
  pages = {112--124},
  doi = {10.1007/978-3-642-36315-3_9},
  editorb = {Hutchison, David and Kanade, Takeo and Kittler, Josef and Kleinberg, Jon M. and Mattern, Friedemann and Mitchell, John C. and Naor, Moni and Nierstrasz, Oscar and Pandu Rangan, C. and Steffen, Bernhard and Sudan, Madhu and Terzopoulos, Demetri and Tygar, Doug and Vardi, Moshe Y. and Weikum, Gerhard},
  editorbtype = {redactor},
  eventtitle = {rc\_conf}
}

@inproceedings{schneiderSATEncodingOptimal2023,
  title = {A {{SAT}} encoding for optimal {{Clifford}} circuit synthesis},
  booktitle = aspdac,
  author = {Schneider, Sarah and Burgholzer, Lukas and Wille, Robert},
  date = {2023},
  eventtitle = {aspdac}
}

@article{shorPolynomialtimeAlgorithmsPrime1997,
  title = {Polynomial-time algorithms for prime factorization and discrete logarithms on a quantum computer},
  author = {Shor, Peter W.},
  date = {1997},
  journaltitle = {SIAM J. Comput.}
}

@article{sivarajahKetRetargetableCompiler2021,
  title = {t|ket⟩: a retargetable compiler for {{NISQ}} devices},
  shorttitle = {t|ket⟩},
  author = {Sivarajah, Seyon and Dilkes, Silas and Cowtan, Alexander and Simmons, Will and Edgington, Alec and Duncan, Ross},
  date = {2021-01-01},
  journaltitle = {Quantum Sci. Technol.},
  volume = {6},
  number = {1},
  pages = {014003},
  doi = {10.1088/2058-9565/ab8e92}
}

@software{treinishQiskitQiskit2022,
  title = {Qiskit/qiskit},
  shorttitle = {Qiskit/qiskit},
  author = {Treinish, Matthew and Gambetta, Jay and Nation, Paul and Qiskit-Bot and Kassebaum, Paul and Rodríguez, Diego M. and De La Puente González, Salvador and Lishman, Jake and Shaohan Hu and Krsulich, Kevin and Garrison, Jim and Bello, Luciano and Yu, Jessie and Marques, Manoel and Gacon, Julien and McKay, David and Gomez, Juan and Capelluto, Lauren and Travis-S-IBM and Mitchell, Abby and Panigrahi, Ashish and Lerongil and Rafey Iqbal Rahman and Wood, Steve and Toshinari Itoko and Pozas-Kerstjens, Alex and Wood, Christopher J. and Divyanshu Singh and Risinger, Drew and Arbel, Eli},
  date = {2022-12-08},
  doi = {10.5281/ZENODO.2573505},
  organization = {{Zenodo}}
}

@article{vidalUniversalQuantumCircuit2004,
  title = {Universal quantum circuit for two-qubit transformations with three controlled-{{NOT}} gates},
  author = {Vidal, G. and Dawson, C. M.},
  date = {2004-01-08},
  journaltitle = {Phys. Rev. A},
  volume = {69},
  number = {1},
  pages = {010301},
  doi = {10.1103/PhysRevA.69.010301},
  langid = {english}
}

@inproceedings{willeMappingQuantumCircuits2019,
  title = {Mapping quantum circuits to {{IBM QX}} architectures using the minimal number of {{SWAP}} and {{H}} operations},
  booktitle = dac,
  author = {Wille, Robert and Burgholzer, Lukas and Zulehner, Alwin},
  date = {2019-06-02},
  eventtitle = {dac}
}

@inproceedings{younisQuantumCircuitOptimization2022,
  title = {Quantum circuit optimization and transpilation via parameterized circuit instantiation},
  booktitle = {qce},
  author = {Younis, Ed and Iancu, Costin},
  date = {2022-09},
  pages = {465--475},
  publisher = {{IEEE}},
  location = {{Broomfield, CO, USA}},
  doi = {10.1109/QCE53715.2022.00068},
  eventtitle = {qce}
}

@inproceedings{zulehnerCompilingSUQuantum2019,
  title = {Compiling {{SU}}(4) quantum circuits to {{IBM QX}} architectures},
  booktitle = aspdac,
  author = {Zulehner, Alwin and Wille, Robert},
  date = {2019},
  pages = {185--190},
  doi = {10.1145/3287624.3287704},
  eventtitle = {aspdac}
}

@article{zulehnerEfficientMethodologyMapping2019,
  title = {An efficient methodology for mapping quantum circuits to the {{IBM QX}} architectures},
  author = {Zulehner, Alwin and Paler, Alexandru and Wille, Robert},
  date = {2019},
  journaltitle = tcad
}
\end{document}